\documentstyle[aps,prb]{revtex}
\begin{document}
\input epsf.sty
\twocolumn[
\hsize\textwidth\columnwidth\hsize\csname %
@twocolumnfalse\endcsname
\draft
\widetext


\title{Vertical Boundary at $x \approx 0.11$ in the Structural Phase Diagram of the
La$_{1-x}$Sr$_{x}$MnO$_{3}$ System $(0.08 \le x \le 0.125)$}
 
\author{D. E. Cox, T. Iglesias,$^{\dagger}$ and E. Moshopoulou$^{\ddagger}$}
\address{Department of Physics, Brookhaven National Laboratory, Upton, NY 11973-5000}

\author{K. Hirota, K. Takahashi, and Y. Endoh$^{*}$}
\address{CREST, Department of Physics, Tohoku University, Sendai 980-8578, Japan}

\date{\today}

\maketitle

\begin{abstract}
The structural phase diagram of the La$_{1-x}$Sr$_{x}$MnO$_{3}$ system in
the compositional range $0.08 \le x \le 0.125$ has been investigated by high-resolution
synchrotron x-ray powder diffraction techniques between
$20-600$~K. Recent studies have reported that there is an unusual rentrant-type phase
transition in this range involving an abrupt change in lattice parameters but no change in
the crystal symmetry, which remains orthorhombic $Pbnm$. The transition to the reentrant
phase is from a ferromagnetic metallic to a ferromagnetic insulating phase with some 
unusual properties.  Our results demonstrate that for samples with $x =0.11-0.125$ there
exist two lower-symmetry structural regions having monoclinic and triclinic symmetry
respectively. There is a sharp first-order transition from the monoclinic to the triclinic
phase coinciding with the  transition to the ferromagnetic insulating phase, and an abrupt
crossover from the orthorhombic $Pbnm$ region with a near-vertical phase boundary just
below $x = 0.11$. The new phases indicate the presence of some novel type of orbital
ordering unlike that found in LaMnO$_{3}$.
\end{abstract} 
 

\phantom{.}
]
\narrowtext
 
\section{INTRODUCTION}
\label{INTRO}

Since the recent discovery of colossal magnetoresistance in doped rare-earth manganate
perovskites of the type La$_{1-x}$A$_{x}$MnO$_{3}$ (${\rm A=Ca,Sr,Ba}$),\cite{Jin} there
has been an enormous amount of attention focussed on the relationships among magnetic,
electrical and structural properties, especially with respect to the delicate and complex
balance between charge, orbital and magnetic ordering. Considerable progress has been made
in understanding the physics of these materials, and it is now widely accepted that
although the traditional double-exchange mechanism originally proposed by
Zener\cite{Zen,And,deG} is qualitatively able to account for the ferromagnetic metallic
(FMM) state observed at intermediate doping levels of $x \approx 0.2-0.4$,\cite{Jon,Van} it
is not capable of explaining many of the other magnetic and transport
properties.\cite{Mil1} Some important recent advances have been the recognition of the
importance of electron-phonon coupling effects,\cite{Roder,Mil2} calculations which show a
tendency for these materials towards phase separation,\cite{Yun,Mor} and the application
of x-ray resonant scattering techniques as a probe of orbital ordering.\cite{Mur} In the
case of LaMnO$_{3}$, such techniques have provided direct evidence of the orbital ordering
of distorted MnO$_{6}$ octahedra associated with a cooperative Jahn-Teller (JT) effect previously
inferred on the basis of bond distances obtained from structure determinations.

One of the important and intriguing remaining questions is the nature of the ferromagnetic
insulating (FMI) phase in lightly-doped La$_{1-x}$Sr$_{x}$MnO$_{3}$ with $x \approx
0.10-0.15$.\cite{Uru}  An usual feature of this phase is that neutron diffraction
measurements have shown that for $x=0.125$ there is first a transition to a FMM phase from
the paramagnetic insulating (PMI) state at higher temperatures, followed by an abrupt
first-order transition to the FMI phase\cite{Kaw1,Kaw}, accompanied by the appearance of a
weak antiferromagnetic component with the A-type magnetic structure similar to that found
in LaMnO$_{3}$.\cite{Wol}  In another neutron diffraction study of crystals with $x = 0.10$
and 0.15, Yamada {\it et al.}\cite{Yam} observed that the FMI phase gave weak superlattice
peaks which they attributed to long-range ordering of Mn$^{4+}$ ions over 1/8 of the Mn
sites, and this type of charge-ordering scenario has been accepted in most subsequent
publications in the literature. However, a recent neutron and synchrotron x-ray study of a
crystal with $x = 0.12$ has demonstrated that conventional charge-ordering does not occur,
and that the key parameter in the transition to the FMI phase is the orbital degree of
freedom of the $e_{g}$ electrons.\cite{End,Ina}

The present paper describes the results of a high-resolution synchrotron x-ray powder
diffraction study carried out on several compositions in the range $x = 0.08-0.125$. The
samples were obtained from fragments of crystals grown in exactly the same way as the one
used by Endoh {\it et al.}\cite{End} The results demonstrate the existence of two
previously unreported lower-symmetry phases with monoclinic and triclinic symmetry
respectively in this region of the phase diagram, separated by a sharp boundary at $x \approx
0.11$ from the JT-ordered LaMnO$_{3}$-type orthorhombic phase which exists at lower values
of $x$. 

\section{STRUCTURAL CONSIDERATIONS}
\label{STRUCTURE}

Before describing the results of the present study, we first review briefly some of the
relevant structural literature. The room-temperature structure of the parent compound
LaMnO$_{3}$ has been reported by several groups to have a distorted perovskite-like
structure with orthorhombic $Pbnm$ symmetry, and lattice parameters $a \approx \sqrt{2}\times
a_{0}$, $b \approx \sqrt{2}\times a_{0}$, $c \approx
2\times a_{0}$, where $a_{0}$ is the parameter of the ideal cubic
lattice\cite{Ele,Norb,Hua,Rod}. As described in a detailed neutron powder diffraction
study by Rodriguez-Carvajal {\it et al.},\cite{Rod} the structure has lattice parameters $a
= 5.5367$, $b = 5.7473$, $c = 7.6929$~\AA , and is characterized by an octahedral tilt
arrangement of the type a$^-$a$^-$c$^+$\cite{Gla} and antiferrodistortive orbital ordering
in the $ab$ plane associated with a cooperative JT effect, resulting in two long Mn-O
bond distances of 2.18 \AA\ and four shorter ones ($2\times 1.91$ and $2\times 1.97~$\AA).
Because of the orbital ordering, the spontaneous orthorhombic strain
$s$ in the basal-plane (defined as $s = 2(a-b)/(a+b)$) is unusually large (about
$-3.7$~\%) in LaMnO$_{3}$ compared to that in most $Pbnm$-type perovskites, and the $c$
axis is significantly shortened ($c/\sqrt{2} = 5.4397$~\AA ). The orbital ordering
disappears at $\sim 750$~K, above which the lattice is metrically cubic. However, Rietveld
refinements show that the octahedral tilt angle decreases very little through the transition, 
and the structure is still unequivocally orthorhombic. At $\sim 1010$~K there is a further
transformation to a rhombohedral structure with $R\bar{3}c$ symmetry, but this $R\bar{3}c$
phase will not be of any concern in the present paper. The JT-ordered structure is usually
designated $O'$, and the ``pseudocubic'' disordered structure $O$ or $O^{*}$. The latter
notation will be used hereafter. 

In contrast to the LaMnO$_{3}$ samples used in the other structural studies, which were
prepared by conventional solid-state synthesis, the one used by Rodriguez-Carvajal {\it et
al.}\cite{Rod} was prepared by crushing single-crystal ingots grown by the floating-zone
method. They emphasize that the structural properties are very sensitive to the amount of
Mn$^{4+}$ introduced during synthesis; for example, the spontaneous orthorhombic strain
decreases, the c parameter increases, and the transition temperatures to the pseudocubic
and rhombohedral phases decrease quite rapidly with increasing Mn$^{4+}$ content. 

The influence of oxygen partial pressure and the formation of Mn$^{4+}$ during the
synthesis of ceramic samples was earlier discussed by Mitchell {\it et al.}\cite{Mit} in
an investigation of the room-temperature structural phase diagram of
La$_{1-x}$Sr$_{x}$MnO$_{3+\delta}$. For $x < 0.125$, oxygen partial pressures $p({\rm O}_{2})$ of 
$< 10^{-2}$ are needed to stabilize the
$Pbnm$ phase. These authors also report that LaMnO$_{3}$ prepared under a
$p({\rm O}_{2})$ of $<10^{-3}$ has a monoclinic structure with $P2_{1}/c$ symmetry, but it
seems clear from the lattice parameters and atomic positions that this structure is almost
identical to the orthorhombic one.\cite{com}

The variation of room-temperature lattice parameters in single crystals
of La$_{1-x}$Sr$_{x}$MnO$_{3}$ with increasing  $x$ is nicely illustrated in
Fig.~\ref{Fig:1} of Urushibara {\it et al.},\cite{Uru} which shows a rapid decrease in
spontaneous orthorhombic strain and an increase in $c/\sqrt{2}$ up to $x =
0.1$, a pseusocubic cell at $x = 0.15$, and a transition to the rhombohedral
structure at $x = 0.175$. The electronic phase diagram shows a narrow
FMI region between $x \approx 0.10-0.15$ at temperatures
below $\sim 150-200$~K. Subsequent neutron diffraction studies on samples
with $x = 0.125$\cite{Kaw,Arg,Pin1} revealed a very unusual structural
transition around 150~K from a JT-distorted $O'$ phase to a pseudocubic
structure metrically similar to that at 300 K and designated $O^{*}$ by
Kawano {\it et al.},\cite{Kaw} who described this as a ``reentrant''
transition, and presented schematic structural and magnetic phase diagrams.
A more detailed structural phase diagram was reported by Pinsard {\it et al.},\cite{Pin2}
who found the low-temperature reentrant phase existed only in the narrow range of
composition $0.10 < x < 0.14$.

In the meantime, Yamada {\it et al.}\cite{Yam} reported a neutron diffraction study of
single crystals with $x = 0.10$ and 0.15, in which they noted the appearance of
superlattice peaks in the FMI region which could be indexed in terms
of a doubling of the orthorhombic $c$ axis. They proposed a model structure consisting of
alternating layers of Mn$^{4+}$ ions ordered on 1/4 of the Mn sites interleaved with
layers of orbitally-ordered Mn$^{3+}$ ions similar to those in LaMnO$_{3}$, but did not
provide any quantitative intensity data in support of this model. 

Similar superlattice peaks were observed in a high-energy synchrotron x-ray study of
crystals with $x = 0.125$ and 0.15 \cite{Nie}, and in a recent neutron and synchrotron
x-ray study of a crystal with $x = 0.12$.\cite{End} A striking result of the latter study
was that x-ray resonant scattering techniques revealed the existence of orbital ordering
in the FMI region below 145~K. Although this behavior shows a
resemblance to that reported for LaMnO$_{3}$,\cite{Mur} there is a very significant
difference, namely that a resonant signal was {\em not} observed in the JT-distorted region
above the transition, but {\em only} in the pseudocubic reentrant region. Thus, the nature
of the JT-type order in the intermediate region is clearly different from the
($d_{3x^2-r^2}/d_{3y^2-r^2}$)-type orbital order in LaMnO$_{3}$. Further evidence for such
a difference is provided by the spin-wave dispersion measurements of Hirota {\it et
al.}\cite{Hir}, who found a dimensional ``crossover'' at around $x = 0.1$ from the
two-dimensional state in LaMnO$_{3}$ to a three-dimensional isotropic ferromagnetic state.
Endoh {\it et al.}\cite{End} also note that the superlattice peaks do not show the expected
resonance features characteristic of charge-ordering. Thus although there is clearly some
kind of structural modulation along the $c$ axis, it cannot be attributed to conventional
charge or orbital ordering. Furthermore, magnetization measurements show clearly that the
low-temperature ferromagnetic phase is stabilized by the application of a magnetic
field,\cite{Sen,Uhl,Noj} whereas a charge-ordered structure would be be expected to melt
in high magnetic fields. The experimental features of the transition are well accounted
for by a theory in which the orbital degree of freedom is considered in the presence of
strong electron correlations,\cite{Ish} and Endoh {\it et al.}\cite{End} conclude that it
is this orbital degree of freedom which is the key parameter in the transition.

\section{EXPERIMENTAL DETAILS}
\label{EXP}

Single crystals of La$_{1-x}$Sr$_{x}$MnO$_{3}$ with nominal compositions
$x = 0.08, 0.10, 0.11, 0.12$ and 0.125 were grown in a floating-zone  furnace. For
subsequent powder diffraction studies, particular  attention was given to the preparation
of suitable samples in order  to minimize possible broadening of the peak profiles due to
induced  microstrain or too small a particle size. Small fragments of the crystals were
crushed and lightly ground in an agate mortar under acetone, and the fraction retained
between $325-400$ mesh  sieves ($\sim 38-44$ $\mu$m) was loaded into an 0.2~mm glass
capillary which was then sealed. High-resolution synchrotron x-ray data were collected  on
several different occasions at beamline X7A at the Brookhaven  National Synchrotron Light
Source with a flat Ge(220) analyzer  crystal, set for a wavelength between
$0.7-0.8$~\AA. With this  diffraction geometry, the instrumental resolution over the
2$\theta$  range $5-30^{\circ}$ is $0.005-0.01^{\circ}$ full-width at half-maximum
(FWHM), an order-of-magnitude better than that of a typical laboratory  diffractometer. 
In some cases, extended data sets were obtained with a linear position-sensitive detector,
which has the advantage of giving greatly-improved counting statistics but with lower
resolution (FWHM $\approx 0.03^{\circ}$).

For low-temperature runs the capillary was mounted in a closed-cycle  helium cryostat, and
scans were made over selected angular regions with a step-size of typically
0.005$^{\circ}$. During each counting interval, the  sample was rocked over several
degrees, which is essential to achieve  powder averaging over crystallites of this size.
For runs above  room-temperature, the capillary was mounted in a wire-wound boron  nitride
tube furnace, and rotated continuously at about 1~Hz during  data collection. Data were
collected over regions  corresponding to the pseudocubic (100), (110), (111), (200), (220)
and (222) reflections, and also the orthorhombic (021) region on selected  occasions.
Least-squares fits to the observed peak profiles were made  based on a pseuso-Voigt
function with appropriate corrections for  low-angle asymmetry due to axial
divergence.\cite{Fin} This  procedure provides an accurate description of the peak shapes,
which  is crucial for the identification of coexisting perovskite-like phases  with low
symmetry and similar lattice parameters. 

The crystal symmetry of a distorted perovskite-like structure can be determined uniquely
from the characteristic splitting of the  pseudocubic $(h00)$, $(hh0)$ and $(hhh)$ peaks.
As mentioned above, the parent compound  LaMnO$_{3}$ has orthorhombic $Pbnm$ symmetry, with
unit cell dimensions $a \approx \sqrt{2}\times a_{0}$, $b \approx \sqrt{2}\times a_{0}$, 
$c \approx 2\times a_{0}$, where $a_{0}$ is the edge of the simple cubic cell, and the split
peaks have the indices and relative  intensities listed in Table~\ref{Table:1}. For lower
crystal symmetries, additional  splittings of some of the peaks occur, as will be described
later.  Another feature of the distorted
$Pbnm$ structure is the appearance of weak superlattice peaks resulting from a combination
of octahedral  tilts and shifts of the large A-site cations from the ideal cubic 
positions. The (021) superlattice peak is especially significant in this respect, since it
provides the means to uniquely identify the $b$ parameter (the corresponding (201) 
reflection being forbidden in $Pbnm$ for symmetry reasons) and is well-resolved from the
adjacent fundamental reflections.  It cannot be  overemphasized that for the very small
distortions observed in some of  these materials, extremely high resolution is needed for
a correct  assignment of the crystal symmetry. The task is further complicated by  the
coexistence of two phases in most of the samples.

\section{RESULTS}
\label{RESULTS}

\begin{figure} 
\centerline{\epsfxsize=3.5in\epsfbox{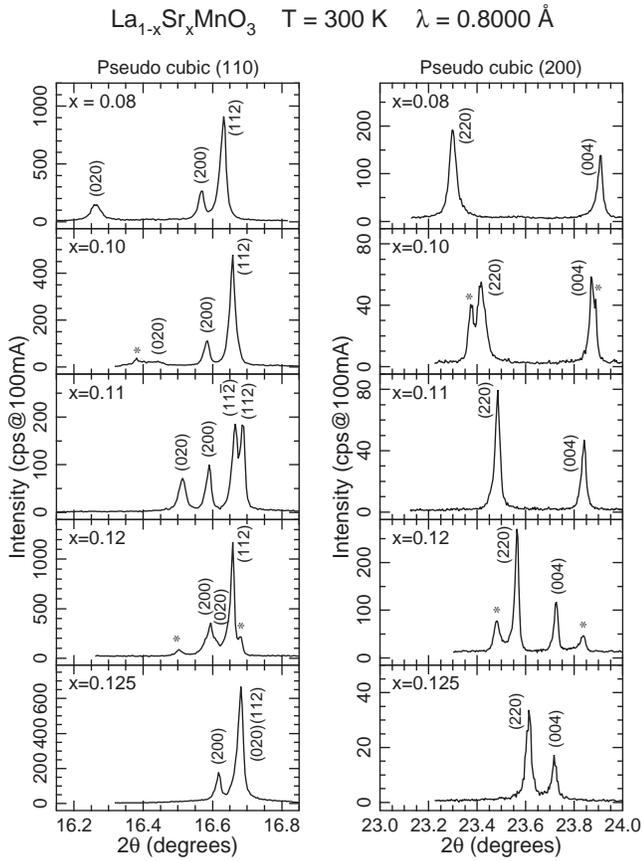}}
\vspace*{0.1in}
\caption{Peak profiles of the pseudocubic (110) and (200) regions at 300 K for the five
nominal compositions $x=0.08,0.10,0.11,0.12$, and 0.125. Peaks from the minority phases are
indicated with asterisks. Note the splitting of the (112) reflection for $x = 0.11$.}
\label{Fig:1}
\end{figure}

Some representative scans at 300 K over the pseudocubic (110) and (200) regions are shown
in Fig.~\ref{Fig:1} for the five compositions. It is  clear from these scans that the
samples with $x = 0.10$ and 0.12 contain  substantial amounts of a second phase with
slightly different lattice  parameters. Furthermore, a careful analysis of the pseudocubic (200) peak
profiles  reveals that there is also a second minority phase in the samples with 
$x = 0.08$ and 0.11, and that only the $x = 0.125$ sample can be considered  to be single
phase. However, the most striking feature in Fig.~\ref{Fig:1} is the splitting at the
orthorhombic (112) position for $x = 0.11$ into two  roughly equal components, which is
conclusive evidence of a lower  symmetry (Table~\ref{Table:1}). Analysis of the complete
set of peak profiles for this  material revealed the true symmetry to be monoclinic. An
approximate  estimate of the phase fractions in the two-phase samples was obtained  from
the relative intensities of the two sets of peaks and  consideration of the predicted
intensity ratios in Table~\ref{Table:1}.

\begin{figure} 
\centerline{\epsfxsize=3.5in\epsfbox{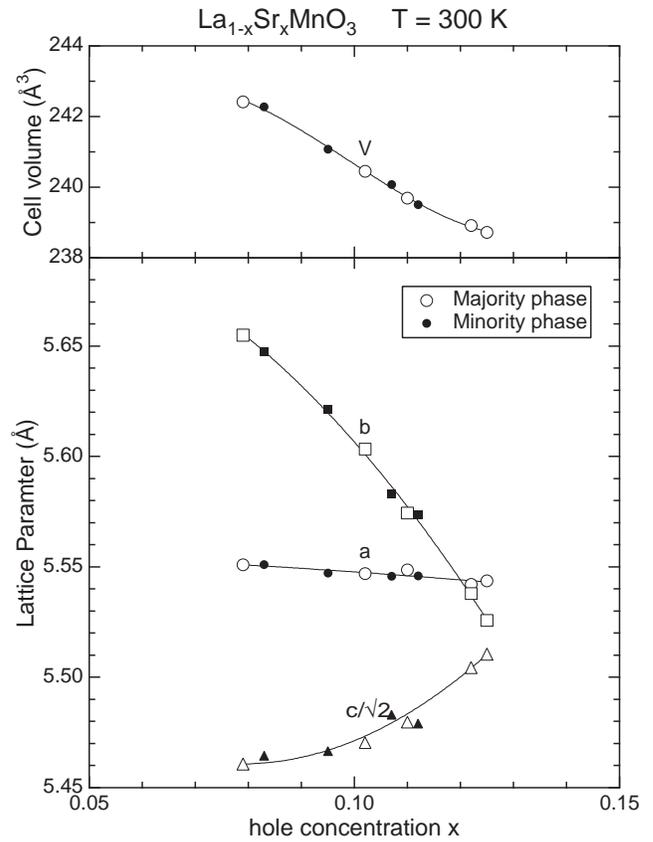}}
\vspace*{0.1in}
\caption{The variation of lattice parameters for the five nominal compositions $x = 0.08,
0.10, 0.11, 0.12$, and 0.125. The estimated ``x-ray'' compositions are derived from the
unit-cell volumes based on the assumption that the actual overall composition is the nominal
one.}
\label{Fig:2}
\end{figure}

The lattice parameters at 300~K were determined from least-squares fits to the peak
positions for each of the five samples, including the  minority phases. If the nominal
compositions of the samples are assumed to be correct, a fairly accurate estimate of the
``true'' compositions  in the two-phase samples can be obtained from the estimated phase 
fractions and either the variation of the $b$ parameter or the unit-cell  volume, both of
which are quite sensitive to small changes in $x$.  The coexistence of two phases in
crystals grown by the floating-zone  technique can be attributed either to
differences in oxygen  content or to fluctuations in Sr concentration in different parts
of boule.  In either case, $x$ should be interpreted as the  Mn$^{4+}$ content or hole
concentration, but only on a relative scale,  since the actual oxygen stoichiometry of the
as-grown sample is not  known precisely. The lattice parameters and compositions derived
from  the above procedure are summarized in Table~\ref{Table:2} and plotted as a function
of $x$ in Fig.~\ref{Fig:2}. These results are qualitatively similar to those  reported by
Urushibara {\it et al},\cite{Uru} i.e. $b$ decreases sharply, $c$ increases fairly
rapidly, and $a$ decreases slowly with increasing
$x$. It is interesting to note that there is a ``cross-over'' at $\sim 300$~K for the
$a$ and $b$ parameters at $x \approx 0.12$, signifying a change of sign in the  basal-plane
strain $s=2(a-b)/(a+b)$, corresponding to the disappearance of JT-type orbital ordering.
The rapid change of $b$ as a function of $x$ is reflected in a tendency for the (020) peak
to be somewhat broadened with respect to the others, as can be seen in Fig.~\ref{Fig:1} for
$x = 0.08$ and 0.10, for example.  From the measured peak-widths it is possible to estimate
the compositional  fluctuation, $\Delta x$, to be $0.001-0.002$ (FWHM) after correction for
instrumental resolution. 

Several sets of measurements were carried out as a function of  temperature in order to
map out the features of the structural transition. For nominal $x = 0.08$, JT-type orbital
ordering  transitions occurred for the majority ($x = 0.079$) and minority  ($x = 0.083$)
phases at $\sim 550$ and $\sim 530$ K respectively, as inferred by a sharp downturn in the
$c$ parameter.  No indication of  any reentrant behavior was observed down to 20 K. The
variation of the lattice parameters of the $x = 0.079$ phase was qualitatively similar to
that of LaMnO$_{3}$,\cite{Rod} but the maximum strain $s$ was much less, about  $-2.0$~\%
vs.\ $-3.7$~\%. Similar behavior was found for nominal $x = 0.10$, in  which orbital
ordering transitions were observed for the majority  ($x = 0.102$) and minority ($x =
0.095$) phases at $\sim 435$ and $\sim 455$ K  respectively, with values of $s$ of
$-1.2$~\% and $-1.5$~\% respectively at 20~K.

\begin{figure} 
\centerline{\epsfxsize=3.5in\epsfbox{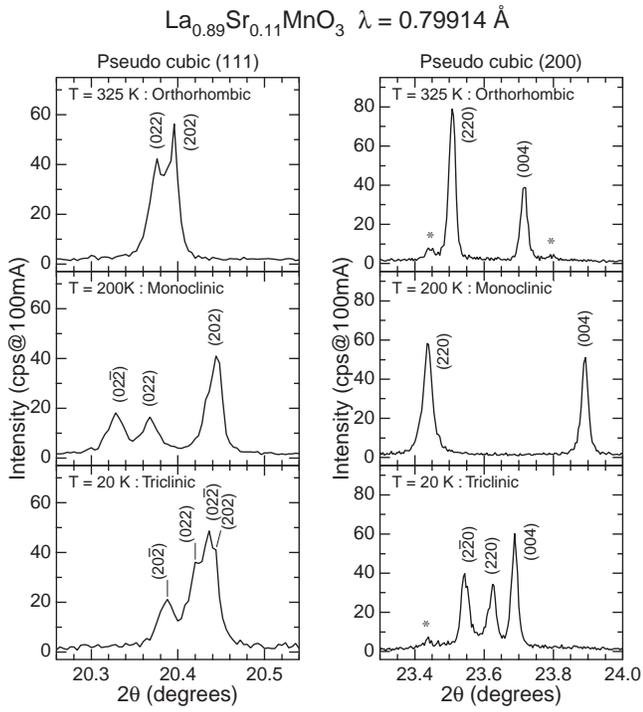}}
\vspace*{0.1in}
\caption{Peak profiles of the pseudocubic (111) and (200) scans for the nominal
composition $x=0.11$ at three temperatures, 325, 200 and 20~K, showing the evolution of the
monoclinic and triclinic phases from the orthorhombic one.}
\label{Fig:3}
\end{figure}

The behavior of the nominal $x = 0.11$ sample was strikingly different,  as illustrated by
the scans of the pseudocubic (111) and (200)  reflections shown in Fig.~\ref{Fig:3}. At 325
K the peak splittings and intensity  ratios confirm that the symmetry is orthorhombic
(Table~\ref{Table:1}). Between  320--105 K, the separation between the (220) and (004) peaks
is  approximately doubled, indicative of a cooperative JT ordered structure in the $ab$
plane,  and the (022) reflection is split into two roughly equal peaks,  consistent with
the monoclinic symmetry noted above. Quite remarkably,  in addition to the abrupt change in
lattice parameters previously  reported\cite{Kaw,Arg,Pin1,Pin2,Uhl} there is another
distortion of the lattice below  105 K, clearly shown by the splitting of the (220) peak
into two  components with about the same intensity. Determination of the crystal  symmetry
is complicated by the large number of peaks and the high degree of overlap, but from a
careful analysis of the peak positions  and intensity ratios it is possible to conclude
that the structure  must be triclinic.

The variation of the lattice parameters and cell volume as a  function of temperature is
shown in Fig.~\ref{Fig:4}. In order to differentiate  between the $a$ and $b$ axes it was
assumed that reflections derived from  the forbidden orthorhombic (201) reflection were
also absent in the  lower symmetry phases.  These assignments are consistent with those
reported in previous studies of $x=0.125$.\cite{Arg,Pin2}  The orthorhombic-to-monoclinic
phase  transformation occurs over a coexistence range of about 5 K centered  around 320 K
(Fig.~\ref{Fig:4}), at which point the two phases are present in  roughly equal
proportions, and there are small but abrupt changes in $b$,
$c$ and $\alpha$ respectively, indicative of a weakly first-order transition. In the
coexistence region there are small changes in $b$ and $c$  for both phases which can be
attributed to compositional fluctuations  with $\Delta x \approx 0.002$. No hysteresis was
observed within an experimental accuracy of $\sim 1$~K. Between $320-150$~K, there is a
smooth increase in $b$ and $a$, and a decrease in $c$ indicative of  a cooperative JT
ordering, but the  orthorhombic strain $s$ is much less than in LaMnO$_{3}$.  Below 150 K
there is a slight downturn in $b$ and upturn in $c$, followed by an abrupt change at $\sim
105$~K associated with the reentrant-type transition from  monoclinic to triclinic symmetry.
There is coexistence of the two  phases over $\sim 5$~K, with a hysteresis on heating of
$\sim 2-3$~K. The behavior  shown in Fig.~\ref{Fig:4} is qualitatively similar to that
previously reported  for samples with $x = 0.125$,\cite{Arg,Pin2} but the reentrant-type
transition is much sharper in the present sample.

It is interesting to note the behavior of the very weak (220) peak from the minority phase
with $ = 0.107$, which is well-resolved at  325 K and below 105 K (as shown by asterisks
in Fig.~\ref{Fig:3}), and just discernible as a small shoulder in between. There is very
little change in the position of this peak  over the whole temperature range, and it is
accordingly clear that this phase does not undergo a reetrant transition.

The sample with nominal $x = 0.12$ shows well-resolved peaks from two phases at 300~K in
the approximate ratio 3:1 (Fig.~\ref{Fig:1}). In this case,  the majority phase with
estimated $x = 0.122$ transforms to an JT-ordered monoclinic phase at $\sim 295$~K
and undergoes a reentrant transition at $\sim 140$~K.  The behavior of the minority phase,
with an estimated $x = 0.112$, is essentially the  same as that of the $x = 0.11$ sample,
with transitions at $\sim 320$~K and  $\sim 120$~K.

Finally, the nominal $x = 0.125$ sample, which appears to be single  phase at 300 K,
undergoes a first-order transition from orthorhombic  to monoclinic at $\sim 280$~K with a
narrow range of two-phase coexistence.  Between $180-140$~K, there is a fairly gradual
approach to a reentrant  transition with a complex two-phase coexistence region, followed
by an  abrupt jump into a single-phase reentrant region. However, in this case the only
indication of lower symmetry is a slight broadening of the  (220) peak.  The overall
variation of the lattice parameters is in good agreement with the results of Pinsard {\it
et al.}\cite{Pin2}; in particular, the unit-cell volumes at 300~K agree very closely,
indicating that the hole concentrations $x$ are the same within 0.002 of each other (see
Fig.~\ref{Fig:2}).

Values of the lattice parameters are listed in Table~\ref{Table:3} for the  majority phase
in each of the five samples at selected temperatures in the different symmetry regions,
designated by $O^{*}$ (pseudocubic), $O'_{JT}$ (JT-ordered orthorhombic), $M_{JT}$
(JT-ordered  monoclinic) and $T$ (triclinic). Also tabulated are the 
basal-plane strain, $s$, which is seen to be positive in the $O^{*}$  and $T$
regions, but negative in the $O'_{JT}$ and $M_{JT}$ regions. The assignment of triclinic
symmetry to the $x = 0.125$ sample at 75~K must be regarded as very tentative, since it
relies mainly on the  decomposition of the (220) profile into two peaks.

\begin{figure} 
\centerline{\epsfxsize=3.5in\epsfbox{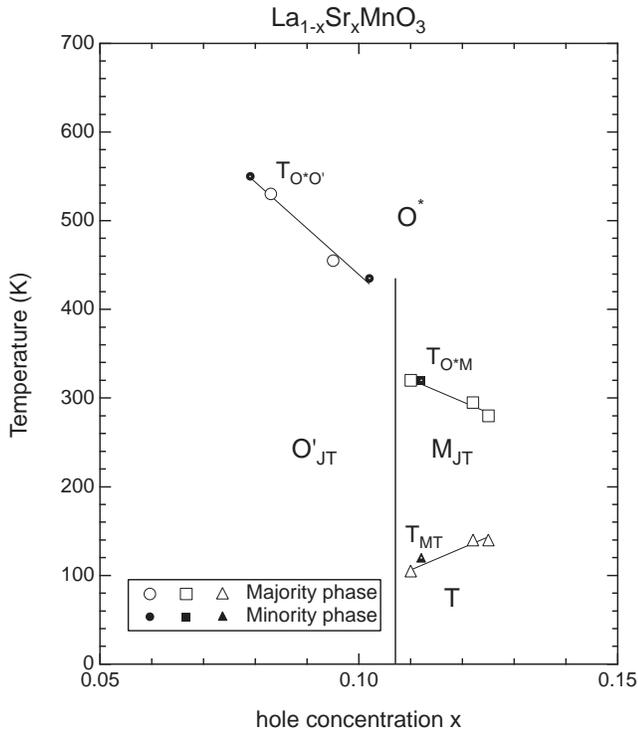}}
\vspace*{0.1in}
\caption{A schematic plot of the phase diagram as a function of temperature for the
``x-ray'' compositions from the samples with nominal $x=0.08,0.10,0.11,0.12$, and 0.125.}
\label{Fig:5}
\end{figure}

Based on the temperature dependence of the lattice parameters of  both the majority and
minority phases, it is possible to construct a revised structural phase diagram in this
narrow region of  composition, as illustrated in Fig.~\ref{Fig:5}. This new phase diagram
contains  the essential features of the ones proposed by Kawano {\it et al.}\cite{Kaw}
and  Pinsard {\it et al.}\cite{Pin2}; in particular the existence of a narrow ``reentrant''
region between $x \approx 0.11-0.13$ at low temperatures, associated with a transition from an
JT-ordered phase to a less-distorted  FMI phase. However, in
striking contrast to the previous results, this region is structurally far more complex
than hitherto supposed: it does not retain orthorhombic symmetry, but  instead undergoes
a sequence of transitions from  orthorhombic-monoclinic-triclinic symmetry. Furthermore,
the line separating the $O'_{JT}$ and $M_{JT}$ regions is extremely sharp, lying  within a
narrow region of hole concentration between $x = 0.107-0.112$. In this region, there is
also a distinct discontinuity between the  rapidly-decreasing slope of the $T_{O^{*}O'}$
and the more gradually decreasing slope of
$T_{O^{*}M}$.  There is an indication of similar  behavior in the data reported by Pinsard
{\it et al.},\cite{Pin2} although their 
$T_{O^{*}O'}$ values are significantly lower than those determined here. This could be due
to small differences  in hole concentration, which for lower values of $x$ is very
sensitive to the final heat treatment and gas atmosphere.\cite{Mit} In this context, we
note that after heating one sample of $x = 0.10$  to 550 K, the initial orthorhombic phase
was not recovered on cooling, but instead a monoclinic phase. Inspection revealed that the
capillary  had not been properly sealed, from which we infer that some oxidation  in air
must have occurred even at this low temperature.

\section{DISCUSSION}
\label{DISCUSSION}

The experiments described above have given a clear description of the different types of
structure in terms of crystal symmetry, but a  detailed determination of the corresponding
atomic shifts in the  monoclinic and triclinic phases is a far more difficult challenge.
Extended data sets were collected with the linear PSD in these regions from the $x = 0.11$
sample in order to check for any weak superlattice peaks associated with the lowering of
symmetry, but no extra peaks were observed at intensity levels of $\sim 0.1$~\% of those of
the strong peaks. In a recent powder diffraction study of charge-ordering in
La$_{0.5}$Ca$_{0.5}$MnO$_{3}$, several superlattice peaks were found at this intensity level
\cite{Rad}, but even if charge-ordering were to occur in the present samples one would
anticipate that any additional peaks would be an order-of-magnitude weaker than for the
latter compound due to the much smaller fraction of Mn$^{4+}$ (1/8 vs.\ 1/2).

\begin{figure} 
\centerline{\epsfxsize=3.5in\epsfbox{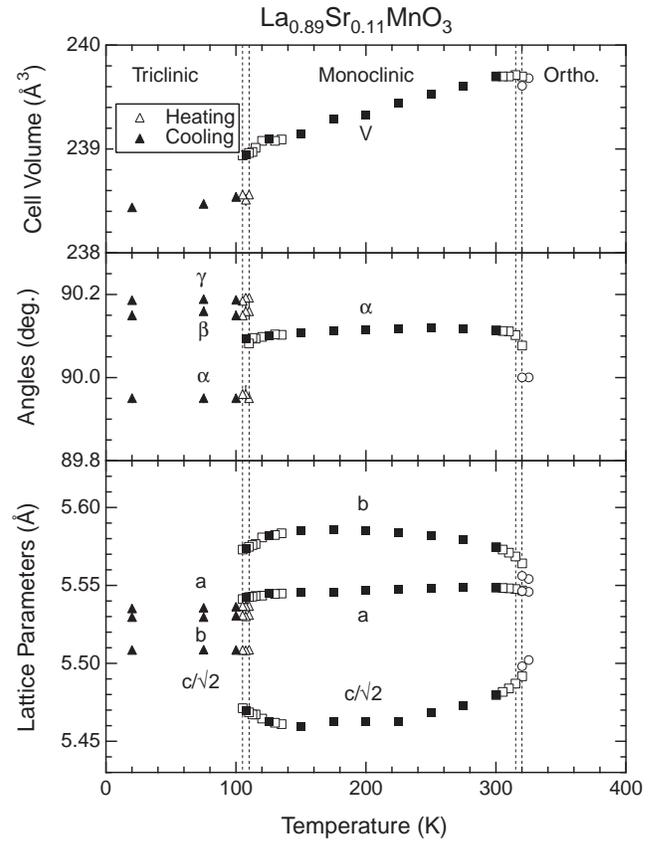}}
\vspace*{0.1in}
\caption{Temperature dependence of (top) cell volume, (middle) angles, and (bottom)
lattice parameters for the nominal composition $x=0.11$.  The vertical dashed lines indicate
the two-phase coexistence regions.}
\label{Fig:4}
\end{figure}

At this stage, we can only speculate about possible features of the  structures which are
consistent with the crystal symmetry and variation of the lattice parameters.  In this
context, it is important to note that pulsed neutron studies in this composition range have
shown that a local JT distortion is present even when the long-range rystallographic
structure shows no such distortion.\cite{Lou}.  What is quite clear from the present
experiments is that there must be fundamental differences between the conventional
JT-distorted structure in the orthorhombic $O'_{JT}$ region, the JT-distorted structure
in the monoclinic $M_{JT}$ region, and the orbitally-ordered triclinic structure in the FMI
region. Although the behavior of the lattice parameters for $x = 0.11$ in the $M_{JT}$ region
shown in Fig.~\ref{Fig:4} implies some kind of JT-ordering in the $ab$ plane, the fact that
there is a monoclinic distortion and the results of the x-ray resonant scattering experiments
both demonstrate that this cannot be the usual C-type antiferrodistortive arrangement found
in the $O'_{JT}$ region.  The fairly abrupt crossover from the latter region to the
$M_{JT}/T$ region at $x \approx 0.1$, which may be related to the dimensional crossover in the
magnetic excitations noted by Hirota {\it et al.}\cite{Hir}, is further evidence of a quite
different type of orbital order.

One possible structural picture is suggested by the nature of the monoclinic distortion,
namely that the unique axis is along [100] in the $Pbnm$ reference frame ($\alpha\neq
90^{\circ}$, so that the $b$ glide-plane symmetry element is preserved while the $n$ and $m$
elements are lost. The loss of the mirror plane would be consistent with a change in JT order
from C-type to G-type, in which there is antiferrodistortive coupling of MnO$_{6}$ octahedra
along the $c$ direction instead of parallel coupling. 
The corresponding space group $P2_1/c$ ($P2_1/b11$ in our non-standard setting) would now
permit reflections of the type $(h0l)$ with $h+l={\rm odd}$, which are forbidden in space
group $Pbnm$. Although no peaks of this type can be detected in the diffraction pattern,
intensity calculations suggest that such peaks would be very weak as long as there is no
change in the $a^{-}a^{-}c^{+}$ octahedral-tilt system characteristic of $Pbnm$ symmetry. 
Note that the G-type JT order in the $M_{JT}$ region is consistent with the {\em
disappearance} of the C-type orbital order as inferred from the missing of resonant x-ray
scattering at $(00l)$ with $l={\rm odd}$ of La$_{0.88}$Sr$_{0.12}$MnO$_{3}$.\cite{End}   
It is natural to assume that the G-type JT order is associated with the G-type orbital order,
which should have resonant scattering at (101) and (011) instead of (001).
Since (011) is structurally forbidden in $P2_1/b11$, while (101) is permitted, we would be
able to see the resonant scattering if the G-type orbital order is realized in the $M_{JT}$
region.
This structure is also related to the ordered-polaron model for the FMI phase proposed in
Fig.~\ref{Fig:4} of Yamada {\it et al.}\cite{Yam}, to the extent that it corresponds to the
configuration in the first and third layers along the doubled $c$ axis, with the
charge-ordered second and fourth layers removed.
Direct confirmation of such a structure from powder data would most likely require a
combination of high-resolution and counting statistics that only an undulator beamline could
provide, and may in fact require a very careful single-crystal study.

It is even more difficult to speculate about the nature of the triclinic structure.
Clearly there is a drastic change in the type of orbital order in which the $b$ glide-plane
is lost and presumably only the inversion center is retained. The doubling of the $c$ axis
observed in single-crystal studies must also be taken into account. The behavior of the
lattice parameters implies a much more complex type of JT orbital order which is no longer
confined to the $ab$ plane, possibly associated with  orbital hybridization of the type
proposed by Endoh {\it et al.}\cite{End}. Elucidation of this structure will certainly have
to await a detailed single-crystal study.

\acknowledgments 

Work at Brookhaven was supported by the U.S. Department of Energy, Division of Materials
Sciences, under contract No.\ DE-AC02-98CH10886.  The National Synchrotron Light Source
is supported by the U.S. Department of Energy, Divisions of Materials
Sciences and Chemical Sciences.  This work was partly supported by a Grant-in-Aid for
Scientific Research from the Ministry of Education, Science, Sports and Culture, by Core 
Research for Evolutional Science and Technology (CREST) from Japan Science 
and Technology Corporation

\vspace{0.5in}

\noindent
$^{\dagger}$Present address: Lucent Technologies, P.I. Tres Cantos, s/n - Zona Oeste,
28760 Tres Cantos, Madrid, Spain.

\noindent
$^{\ddagger}$Present address: Institute for Materials Science, National Center for
Scientific Research ``Demokritos'', 15310 Agia Paraskevi, Athens, Greece.

\noindent
$^{*}$Present address: CREST, Institute for Materials Reserach, Tohoku
University, Katahira 2-1, Aoba-ku, Sendai 980-8577, Japan.


\newpage

\begin{table*} 
\caption{Indicies of split peaks derived from a cubic perovskite cell with
lattice parameter $a_{0}$ in a distorted cell with $a \approx
\protect\sqrt{2} \times a_{0}$, $b \approx \protect\sqrt{2} \times a_{0}$, $c
\approx 2\times a_{0}$, as the symmetry is successively lowered to
orthorhombic, monoclinic and triclinic.  The relative intensities within
each group are approximately equal to the multiplicities of the powder
reflections given in parentheses.}
\vspace*{0.1in}
\begin{tabular}{ccccccc}
Cubic & & Orthorhombic & & Monoclinic & & Triclinic\\
      & & $(\alpha,\beta,\gamma=90^{\circ})$
      & & $(\alpha \ne 90^{\circ})$
      & & $(\alpha,\beta,\gamma \ne 90^{\circ})$\\
\tableline
& &
\parbox{0.5in}{110 (4)} &
$-$ &
\parbox{0.5in}{110 (4)} &
$<$ &
\parbox{0.5in}{110 (2) \\ 1$\bar{1}$0 (2)}
\\
\parbox{0.5in}{100 (6)} &
$<$ &
\\
& &
\parbox{0.5in}{002 (2)} &
$-$ &
\parbox{0.5in}{002 (2)} &
$-$ &
\parbox{0.5in}{002 (2)}
\\
\tableline  
& &
\parbox{0.5in}{200 (2)} &
$-$ &
\parbox{0.5in}{200 (2)} &
$-$ &
\parbox{0.5in}{200 (2)}
\\ 
& &
\parbox{0.5in}{020 (2)} &
$-$ &
\parbox{0.5in}{020 (2)} &
$-$ &
\parbox{0.5in}{020 (2)}
\\ \\
\parbox{0.5in}{110 (12)} &
$<$ 
& & &
\parbox{0.5in}{112 (4)} &
$<$ &
\parbox{0.5in}{112 (2) \\ 1$\bar{1}$2 (2)}
\\
& &
\parbox{0.5in}{112 (8)} &
$<$
\\
& & & &
\parbox{0.5in}{11$\bar{2}$ (4)} &
$<$ &
\parbox{0.5in}{11$\bar{2}$ (2) \\ 1$\bar{1}\bar{2}$ (2)}
\\
\tableline
& & 
\parbox{0.5in}{202 (4)} &
$-$ &
\parbox{0.5in}{202 (4)} &
$<$ &
\parbox{0.5in}{202 (2) \\ 20$\bar{2}$ (2)}
\\ \\
\parbox{0.5in}{111 (8)} &
$<$ & & &
\parbox{0.5in}{022 (2)} &
$-$ &
\parbox{0.5in}{022 (2)}
\\
& &
\parbox{0.5in}{022 (4)} &
$<$ 
\\
& & & &
\parbox{0.5in}{02$\bar{2}$ (2)} &
$-$ &
\parbox{0.5in}{02$\bar{2}$ (2)}
\\
\end{tabular}
\label{Table:1}
\end{table*}

\clearpage

\begin{table*} 
\widetext
\caption{Lattice parameters and unit cell volumes at 300~K for the
majority and minority phases in La$_{1-x}$Sr$_{x}$MnO$_{3}$ with nominal
compositions $x=0.08,0.10,0.11,0.12$ and $0.125$.  The ratios of the
majority and minority phases were estimated from the relative
intensities of the $(220)/(004)$ pairs of reflections, and the x-ray
compositions were derived from these ratios and the unit-cell volume as
described in the text.  The temperature of the $O^{*}-O'_{JT}$ transition
was chosen to be where the sharpest decrease in the $c$ lattice parameter
occurred.}
\vspace*{0.1in}
\begin{tabular}{llrddddcrc}
$x$ & $x$ & Ratio & $a$  & $b$ & $c$ & $\alpha$ & Vol.\ & $T_{O^{*}O}$ &
$T_{MT}$ \\
     & (x-ray) &     & (\AA) & (\AA) & (\AA) & ($^{\circ}$) & (\AA$^{3}$) & (K) & (K) \\
\tableline
0.08 & 0.079 & 80~\% & 5.5509 & 5.6549 & 7.7226 & 90.0 & 242.41 & 550 & -- \\
     & 0.083 & 20~\% & 5.5510 & 5.6476 & 7.7280 & 90.0 & 242.27 & 530 & -- \\
\tableline
0.10 & 0.102 & 75~\% & 5.5469 & 5.6033 & 7.7362 & 90.0 & 240.45 & 435 & -- \\
     & 0.095 & 25~\% & 5.5472 & 5.6212 & 7.7308 & 90.0 & 241.07 & 455 & -- \\
\tableline
0.11 & 0.110 & 95~\% & 5.5486 & 5.5743 & 7.7494 & 90.114 & 239.69 & 320 & 105 \\
     & 0.107$^{(a)}$ & 5~\% & 5.5457 & 5.5828 & 7.7541 & 90.0 & 240.07 & $>325$ & -- \\
\tableline
0.12 & 0.122$^{(b)}$ & 75~\% & 5.5421 & 5.5379 & 7.7842 & 90.0 & 238.91 & 295 & 140 \\
     & 0.112$^{(b)}$ & 25~\% & 5.5459 & 5.5734 & 7.7485 & 90.120 & 239.50 & 320 & 120 \\
\tableline
0.125 & 0.125 & 100~\% & 5.5437 & 5.5257 & 7.7929 & 90.0 & 238.72 & 280 & 140
\end{tabular}
\vspace*{0.1in}
(a) Estimated from lattice parameters at 325~K.

(b) Lattice parameters from interpolation between 295~K and 305~K.
\label{Table:2}
\narrowtext
\end{table*}


\begin{table*} 
\widetext
\caption{Lattice parameters of the majority phase in each of the five samples studied at
selected tempreatures in the various temperature regions, designated as $O^{*}$
(orthorhombic pseudocubic), $O'_{JT}$ (JT-ordered orthorhombic), $M_{JT}$ (JT-ordered
monoclinic), and $T$ (triclinic).  The assignment of the $a$ and $b$ parameters in the
$M_{JT}$ and $T$ regions is based on the assumption that the forbidden orthorhombic (201)
reflection is also absent in the latter regions.  $s$ is the spontaneous basal
plane strain defined as $2(a-b)/(a+b)$.  The figures in parentheses represent the
estimated standard deviation from the least-squares fits, referred to the least
significant digit(s).}
\vspace*{0.1in}
\begin{tabular}{lrllldl}
$x$ & $T$ (K) & $a$ (\AA) & $b$ (\AA) & $c$ (\AA) & $s$ (\%) & Phase \\
\tableline
0.079 & 580 & 5.5675(6) & 5.5496(6)  & 7.8560(9)  &    0.32 & $O^{*}$   \\
      &  20 & 5.5497(9) & 5.6646(9)  & 7.6845(18) & $-$2.05 & $O'_{JT}$ \\
\tableline
0.102 & 570 & 5.5678(5) & 5.5344(4)  & 7.8387(7)  &    0.60 & $O^{*}$   \\
      &  20 & 5.5534(3) & 5.6212(11) & 7.7108(15) & $-$1.21 & $O'_{JT}$ \\
\tableline
0.110 & 325 & 5.5459(2) & 5.5542(2)  & 7.7811(2)  & $-$0.15 & $O^{*}$   \\
      & 200 & 5.5468(2) & 5.5853(2)  & 7.7251(2)  & $-$0.69 & $M_{JT}$  \\
      &     & $\alpha=90$.115(4) \\
      &  20 & 5.5353(5) & 5.5295(5)  & 7.7903(1)  &    0.10 & $T$       \\
      &     & $\alpha=89$.95(1) & $\beta=90$.15(1) & $\gamma=90$.19(2)  \\
\tableline
0.122 & 305 & 5.5425(5) & 5.5346(5)  & 7.7857(4)  &    0.14 & $O^{*}$   \\
      & 175 & 5.5448(4) & 5.5690(4)  & 7.7243(3)  & $-$0.44 & $M_{JT}$  \\
      &     & $\alpha=90$.103(9) \\
      &  65 & 5.5315(8) & 5.5196(8)  & 7.7893(1)  &    0.22 & $T$       \\
      &     & $\alpha=89$.93(2) & $\beta=90$.18(1) & $\gamma=90$.09(1)  \\
\tableline
0.125 & 300 & 5.5437(6) & 5.5257(6)  & 7.7929(8)  &    0.33 & $O^{*}$   \\
      & 200 & 5.5455(2) & 5.5622(2)  & 7.7333(3)  & $-$0.30 & $M_{JT}$  \\
      &     & $\alpha=90$.134(2) \\
      &  75 & 5.5338(1) & 5.5180(3)  & 7.7879(1)  &    0.29 & $T$       \\
      &     & $\alpha=89$.96(1) & $\beta=90$.14(1) & $\gamma=90$.03(1)  
\end{tabular}

\label{Table:3}
\narrowtext
\end{table*}

\end{document}